\documentclass{emulateapj}

\newcommand{\be}{  \begin{eqnarray} }
\newcommand{\ee}{  \end{eqnarray} }
\newcommand{\msun}{ M_{\odot}}
\newcommand{\mdot}{\dot{M}}
\newcommand{\teff}{T_{\rm eff}}
\newcommand{\tfor}{T_\ast}
\newcommand{\taufor}{\tau_\ast}
\newcommand{\rhofor}{\rho_\ast}
\newcommand{\kappaes}{\kappa_{\rm es}}
\newcommand{\kappaab}{\kappa_{\rm ab}}
\newcommand{\cgas}{c_{\rm g}}
\newcommand{\kbol}{k_{\rm B}}

\shorttitle{Spectral Fitting of Black Hole Binaries}
\shortauthors{Davis et al.}

\begin{document}
\title{Testing Accretion Disk Theory in Black Hole X-ray Binaries}
\author{Shane W. Davis\altaffilmark{1}, Chris Done\altaffilmark{2}, and 
Omer M. Blaes\altaffilmark{1}}
\altaffiltext{1}{Department of Physics, University of California, Santa Barbara,
CA 93106}
\altaffiltext{2}{Department of Physics, University of Durham, South Road,
Durham, DH1 3LE, UK}

\begin{abstract}

We present results from spectral modeling of three black hole X-ray binaries:
LMC X-3, GRO J1655-40, and XTE J1550-564.  Using a sample of disk dominated
observations, we fit the data with a range of spectral models that includes a
simple multitemperature blackbody (DISKBB), a relativistic accretion disk model
based on color-corrected blackbodies (KERRBB), and a relativistic model based
on non-LTE atmosphere models within an $\alpha$ prescription (BHSPEC).  BHSPEC
provides the best fit for a {\it BeppoSAX} observation of LMC X-3, which has
the broadest energy coverage of our sample.  It also provides the best fit for
multiple epochs of {\it Rossi X-ray Timing Explorer} ({\it RXTE}) data in this
source, except at the very highest luminosity ($L/L_{\rm Edd}\gtrsim 0.7$), where 
additional physics must be
coming into play.  BHSPEC is also the best-fit model for multi-epoch {\it RXTE}
observations of GRO J1655-40 and XTE J1550-564, although the best-fit
inclination of the inner disk differs from the binary inclination. All our fits
prefer $\alpha=0.01$ to $\alpha=0.1$, in apparent disagreement with the large
stresses inferred from the rapid rise times observed in outbursts of these two sources.
In all three sources our fits imply moderate black hole spins ($a_\ast\sim 0.1 - 0.8$),
but this is sensitive to the reliability of independent measurements of these
system parameters and to the physical assumptions which underly our spectral models.

\end{abstract}


\section{Introduction}
 \label{intro}

Black hole X-ray binaries (BHBs) are known to occupy distinct spectral states
which can be characterized by the relative contribution of thermal and
non-thermal emission components (e.g. McClintock \& Remillard 2004).  The most
well-understood of these states is the thermal dominant (or high/soft) state.
Here most of the flux is in the thermal component, which is generally assumed to
be emission from a radiatively efficient, geometrically thin accretion disk
(Shakura \& Sunyaev 1973).  The disk is believed to extend deep within the
gravitational field of the black hole, and this makes spectral modeling of
this state an important probe of both the physics of relativistic accretion
disks and the properties of black holes.

In standard treatments of black hole accretion disks, the emitting matter
extends down to the innermost stable circular orbit (ISCO), which is determined
by the mass and spin of the black hole.  Such treatments usually assume a
``no-torque'' inner boundary condition at this radius (e.g. Novikov
\& Thorne 1973), but magnetic fields may in fact exert such torques
(Gammie 1999; Krolik 1999; Hawley \& Krolik 2002), increasing the radiative
efficiency of the disk.  The structure and emission of these disks are therefore
sensitive to the mass and spin of the black hole as well as any torque which may
be present.

This sensitivity makes accretion disk spectral modeling a potential way to
measure or constrain black hole spin (e.g. Ebisawa et al. 1993; Zhang et al. 1997;
Shafee et al. 2006; Middleton et al. 2006).
This method requires a model of the radial profile of
effective temperature in the gravitational field of the black hole, calculation
of the relativistic transfer function from the disk surface to an observer at
infinity (Cunningham 1975), and spectral modeling of the surface emission in the
local rest frame of the disk.  Most implementations of this method approximate
one or all of these components.  One common approximation is to assume that the
disk surface emission is a blackbody or, more generally, a color-corrected
blackbody,
\be 
I_{\nu}=f^{-4}B_{\nu}(fT_{\rm eff}), \label{e:ccbb} 
\ee 
where $I_{\nu}$ is the specific intensity, $T_{\rm eff}$ is the effective
temperature, $B_{\nu}$ is the Planck function, and $f$ is the spectral hardening
factor (or color-correction), typically assumed to be around 1.7 
(Shimura \& Takahara 1995).  One of
the most sophisticated models of this type is the KERRBB model (Li et al. 2005)
for Xspec (Arnaud 1996).  It accounts for the relativistic effects on the disk
effective temperature profile and the relativistic transfer function.  
Potential difficulties with the color-corrected blackbody approximation exist.
The local spectrum may not be well approximated by an isotropic, color-corrected
blackbody due to limb darkening and frequency dependent absorption opacities.  
Even if it can, one must still specify $f$.
It has been suggested that $f$ is a relatively strong function
of accretion rate or of the fraction of energy emitted in a corona (Merloni et
al. 2000).

Relativistic models (Davis et al. 2005) now exist which calculate
the non-LTE vertical disk structure and radiative transfer self-consistently
using the TLUSTY stellar atmospheres code (Hubeny \& Lanz 1995).  The
relativistic effects on photon geodesics are accounted for with ray tracing
methods (Agol 1997).  These spectral models have now been implemented in an
Xspec table model (BHSPEC; Davis \& Hubeny, 2006). By calculating
values of $f$ appropriate for use with KERRBB, these models have already been
used to estimate black hole spins from KERRBB fits to two BHBs: GRO J1655-40
and 4U 1543-47 (Shafee et al. 2006).

In this work, we circumvent the color-corrected blackbody approximation
entirely by fitting the BHSPEC model directly to BHB observations.  Since the
model does not include irradiation of the disk surface, we focus our efforts
on thermal dominant state observations in which the non-thermal emission is a small
fraction of the total flux. Fortunately, a sample of such observations made
with the {\it Rossi X-ray Timing Explorer} ({\it RXTE}) already exists
(Gierlinski \& Done 2004; hereafter GD04).

BHSPEC still assumes that the disk emission extends only down to the ISCO
and then effectively ceases due to rapid in-fall of matter interior to this
radius.  This assumption appears to be consistent with spectral modeling of
BHBs in the thermal dominant state in that the luminosity $L$ is seen to scale roughly
with the fourth power of the color temperature $T_{\rm c}$ in several different
sources (Kubota et al. 2001; Gierlinski \& Done 2004).  This
suggests that as sources vary by over an order of magnitude in luminosity,
there is a roughly constant emitting area and thus a relatively constant inner
radius to the disk.  It is therefore very natural to associate such a stable
inner radius with the ISCO of the black hole, though the `emission edge'
need not coincide exactly with the ISCO (Krolik \& Hawley 2002).

Not all of these sources follow the $L \propto T_{\rm c}^4$ relation exactly,
however.  In several cases a relative hardening is seen 
with increasing $L$ (see e.g. GD04; Kubota \& Makishima 2004; Shafee et al. 2006).
A potential explanation for this hardening is that 
advection may be becoming increasingly important as these sources approach 
the Eddington limit (Kubota \& Makishima 2004).
Alternatively, this behavior is qualitatively consistent with the increased 
spectral hardening with accretion rate in the local disk atmospheres, ignoring 
advection (Davis et al. 2005; Shafee et al. 2006).  In
the BHSPEC model, the precise nature of the hardening depends strongly on the
variation in surface density with radius in the disk.  Currently, the surface
density is determined by assuming the vertically averaged stress is proportional
to the vertically averaged total pressure with a constant of proportionality
$\alpha$.  However, more general stress prescriptions could be implemented in the
future. Therefore, spectral modeling could potentially provide a constraint on
the nature of the stresses in these systems.

In this paper we fit these fully-relativistic, non-LTE accretion disk models to
{\it RXTE} and {\it BeppoSAX} observations of BHBs in the thermal dominant state.
Our purpose is two-fold:  we want to test the applicability of the spectral
models to these observations, and having then found suitable representations
of the data, we use them to infer black hole spins and infer properties of the
stresses in these system.
We review our spectral fitting and results in section \ref{specmod}, discuss the
implications of these results in section \ref{discus}, and summarize our
conclusions in section \ref{conc}.  In the Appendix we develop a simplified model
to understand and motivate the variation of the BHSPEC spectra with accretion rate.

\begin{figure}
%\plotone{f1.eps}
\includegraphics[width=0.46\textwidth]{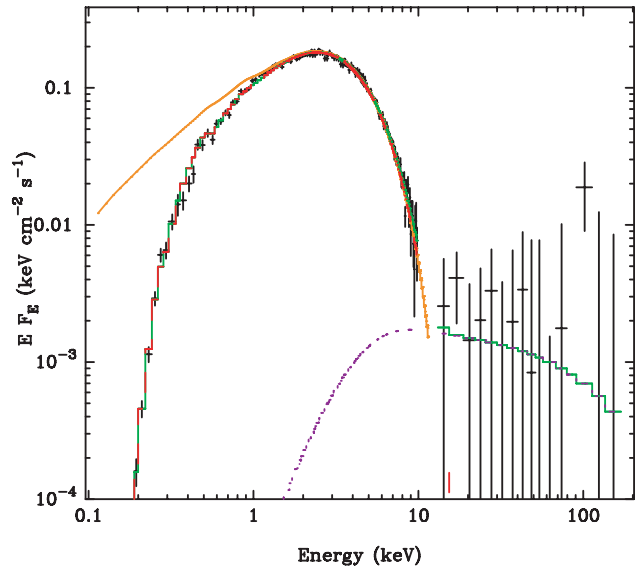}
\caption{The unfolded spectrum for the {\it BeppoSAX}
observation of LMC X-3 using the best fit BHSPEC models for $i=67^{\circ}$, 
$D=52$ kpc, and $\alpha=0.01$.  The total model component (green, solid curve),
BHSPEC (red, long-dashed curve) and COMPTT (violet, short-dashed curve) are
plotted.  The unabsorbed BHSPEC model (orange, solid curve) is also shown.
\label{fig1}}
\end{figure}

\section{Spectral Modeling}
\label{specmod}

Several properties of BHBs make them particularly well suited both for testing
accretion disk theory and for measuring the unknown black hole properties of
interest.  There are several sources for which precise, independent
measurements of the mass of the primary and 
the binary inclination are
available from light curve modeling of the secondary star (e.g. Orosz \& Bailyn
1997). Typically, the distance to BHBs are known with less precision, but there
are exceptions.  As can be seen in Table \ref{tbl1}, the distances to LMC X-3 and
GRO J1655-40 are both claimed to be known to better than 10\%.  
BHBs have an advantage over most active galactic nuclei because of their relatively
short timescales
for large changes in the bolometric luminosity, which allows the same source to
be observed in the same spectral state at appreciably different accretion
rates. The variation of the BHSPEC model with accretion rate provides a precise
quantitative prediction which is sensitive to the assumed stress prescription.  
Therefore, simultaneous fitting of multiple observations of the same source at
different epochs provides a much more powerful constraint and potentially more
information than a single fit to a single epoch.

We compare three accretion disk models for the soft, thermal emission: the
multicolor disk model (DISKBB in Xspec, Mitsuda et al. 1984), KERRBB, and BHSPEC.
DISKBB is the most commonly used spectral model, but it neglects relativity.  
Both KERRBB and BHSPEC include relativistic effects.  In addition, BHSPEC includes 
atmosphere physics which we are interested in testing in this paper.  In particular,
we are interested in examining whether BHSPEC with a fixed $\alpha$ can explain the
variation in the spectral hardening with accretion rate.  As a control, we compare
these results with KERRBB at a fixed $f$.

We also model the neutral absorption along the line of sight, and though we focus
on observations inferred to be disk dominated, we need an additional component to
account for the non-thermal emission which is present. It is widely believed
that the coronal emission is due to inverse Compton scattering of seed photons
from the accretion disk.  In this case, a power law will tend to overestimate the
flux at low energies.  Therefore, we prefer to approximate the non-thermal
emission with the COMPTT Comptonization model (Titarchuk 1994) in our spectral
fits. We specify a disk geometry and fix the high energy cutoff in this model to
50 keV, a value high enough not to affect our fits.  We also tie the seed photon
temperature to the DISKBB model temperature
when DISKBB is used to model the soft emission.  For KERRBB and BHSPEC we fix
this parameter at the best fit DISKBB value.  This leaves two free parameters --
an optical depth and a normalization for each data set.

DISKBB is a relatively simple model with only two parameters: the temperature at
the inner edge of the disk $T_{\rm in}$ and the model normalization. BHSPEC and
KERRBB both share a number of model parameters which need to be specified or fit.
The black hole spin $a_{\ast}\equiv a/M$ and the accretion rate $\mdot$ are free
parameters in all fits unless stated otherwise. For BHSPEC $\mdot$ is
parameterized by $\ell \equiv L/L_{\rm Edd}$ where $L_{\rm Edd}$ is the Eddington
luminosity for completely ionized hydrogen.  This value is converted to an accretion
rate by assuming an efficiency $\eta$ which corresponds to the fraction of
gravitational binding energy at infinity which is converted to radiation. Other
parameters include black hole mass $M$, disk inclination $i$, and distance to the
source $D$. For KERRBB $D$ is an explicit parameter, but for BHSPEC
$D=10/\sqrt{N}$ kpc where $N$ is the model normalization.
In some cases $M$, $i$, and $D$ (or $N$) are fixed at the estimates
given in Table \ref{tbl1}.  Though $M$ is always fixed, there are also cases
where $i$ and $D$ are left as free parameters.  The estimates for $i$ in Table
\ref{tbl1} are all estimates of the binary inclination. However, there is
no guarantee that the angular momentum of the black hole, and therefore the inner
accretion disk, is aligned with the binary (Bardeen \& Petterson 1975). If one
assumes that the jet axis is aligned with the angular momentum vector of the
black hole, then misalignments may be common (Maccarone 2002). XTE J1550-564
and GRO J1655-40 are among the sources for which misalignment can be inferred.  
Therefore, we also consider fits in which $i$ and $D$ are free parameters with
$i$ unconstrained and $D$ allowed to vary within the confidence intervals in
Table \ref{tbl1}.  These models also allow for the presence of a torque on the
inner disk which is parameterized by the increase in efficiency due to the torque
relative to the efficiency of the untorqued disk $\Delta \eta/\eta$.

\begin{deluxetable*}{lccccl}
\tablecolumns{6}
\tablecaption{Source Descriptions\label{tbl1}}
\tablewidth{0pt}
\tablehead{
\colhead{Name} &
\colhead{Mass} &
\colhead{Distance} &
\colhead{Inclination} &
\colhead{$N_{\rm H}$} &
\colhead{Reference}\\
 &
\colhead{$(\msun)$} &
\colhead{(kpc)} &
\colhead{(deg)} &
\colhead{$(10^{22}\,{\rm cm}^{-2})$} &
}

\startdata
LMC X-3 &
$7 (5-11)$ &
$52 (51.4-52.6)$ &
$67 (65-69)$ &
$0.04$ &
1, 2, 3, 4, 5 \\
XTE J1550-564 &
$10 (9.7-11.6)$ &
$5.3 (2.8-7.6)$ &
$72 (70.8-75.4)$ &
$0.65$ &
6, 7 \\
GRO J1655-40 &
$7 (6.8-7.2)$ &
$3.2 (3.0-3.4)$ &
$70 (64-71)$ &
$0.8$ &
8, 9, 10, 11\\
\enddata

\tablerefs{(1) Soria et al. (2001); (2) Cowley et al. (1983); (3) di Benedetto (1997); 
(4) Kuiper et al. (1988); (5) Page et al. (2003); (6) Orosz et al. (2002);
(7) Gierli\'nski \& Done (2003); (8) Shahbaz et al. (1999);
(9) Hjellming \& Rupen (1995); (10) van der Hooft et al. (1998);
(11) Gierli\'nski et al. (2001)}

\end{deluxetable*}

There are other parameters which are not shared between BHSPEC and KERRBB. Two
additional parameters for BHSPEC are $\alpha$ and the metal abundance. In
the BHSPEC model, the stress is given by
\be
\tau_{R\phi}=\alpha P
\label{taurphi}
\ee
where $\tau_{R\phi}$ is the vertically averaged accretion stress and $P$ is the
vertically averaged total pressure.  Usually, the metal abundances are fixed at
the solar value, but we also consider fits with three times solar metallicity.
The color correction parameter $f$ must be chosen for KERRBB, and we fix
$f=1.7$ unless stated otherwise.  The $f$ value
which brings KERRBB into best agreement with BHSPEC is a function of $\ell$
(Shafee et al. 2006), but this choice
makes KERRBB roughly consistent over the range of $\ell$ we consider. We always fix the parameters
{\it rflag} and {\it lflag} so that the spectra are limb darkened and reprocessed
emission from self-irradiation is ignored. Test cases suggest that these choices
do not have a significant effect on the quality of fit or the inferred values for
$a_{\ast}$.

\subsection{Source Selection}
\label{soursel}

Our work is motivated in part by spectral fitting of BHBs performed by Gierlinski
\& Done (2004). They already provided a sample of sources with {\it RXTE}
observations in which a low fraction (under 15\%) of the bolometric flux is
inferred to be in the non-thermal component.  We focus on three of these sources:
LMC X-3, XTE J1550-564 (hereafter J1550), and GRO J1655-40 (hereafter J1655).  
Each one ranges over nearly a decade or more in bolometric luminosity (see Figure
2 of Gierlinski \& Done 2004).  This makes them particularly well suited for
constraining the spectral variation over a range of $\mdot$. The properties
of these sources are summarized in Table \ref{tbl1}.  All have reasonably precise
mass estimates, and the distances to LMC X-3 and J1655 both have relatively small
uncertainties.  The distance to J1550 is less well constrained, but we still
include it in our sample because it spans the widest range of luminosities.

One drawback of {\it RXTE} is its lack of soft X-ray coverage.  The thermal
components of BHBs typically peak at photon energies near or below 1 keV,
whereas the {\it RXTE} band extends down to only $\sim3$ keV. 
Since one of our goals is to
test the applicability of the underlying accretion disk model, we would like to cover
as much of the SED as possible.  Even if we use other observatories, we face the
difficulty that most BHBs lie in or near the Galactic plane and are heavily absorbed by
the interstellar medium along the line-of-sight. Therefore, we also examine a
{\it BeppoSAX} observation of LMC X-3 which has a low line-of-sight absorption
column.  {\it
BeppoSAX} is well-suited for this purpose as it covers a very broad range of
photon energies extending down to a tenth of a keV, but lacks the effective area of
{\it XMM} or {\it Chandra} and their corresponding pile-up problems for such
bright sources.

\begin{figure}
%\plotone{f2.eps}
\includegraphics[width=0.48\textwidth]{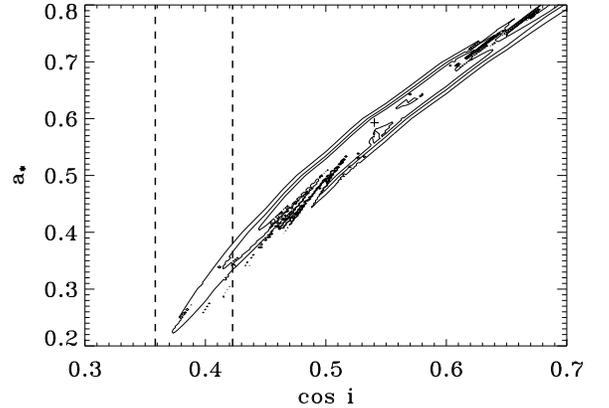}
\caption{The 66\%, 90\%, and 99\% confidence contours in the $a_{\ast}$--$\cos i$
plane for the best fit BHSPEC model
($\alpha=0.01$, $i$ free) to the {\it BeppoSAX} LMC X-3 data.  The 
vertical dashed lines mark uncertainty limits inferred for the binary inclination.
\label{fig2}}
\end{figure}

\subsection{LMC X-3}
\label{lmc}

\subsubsection{{\it BeppoSAX} Data}
\label{lmcsax}

The low energy coverage of {\it BeppoSAX}
makes the fits particularly sensitive to our model for the neutral absorption.  
We therefore include the line-of-sight absorption column as a free parameter in
our fits to the {\it BeppoSAX} data.  For our best fit BHSPEC model we find 
$N_H=5.92^{+0.31}_{-0.27} \times 10^{20}$ cm$^{-2}$.  This is
higher than values inferred from 21 cm absorption ($3.2 \times 10^{20}$
cm$^{-2}$, Nowak et al. 2001) or fits to the neutral oxygen edge in observations
with the Reflection Grating Spectrometer on-board {\it XMM-Newton} ($3.8 \pm 0.8
\times 10^{20}$ cm$^{-2}$, Page et al. 2003), but lower than those found in
previous modeling of the {\it BeppoSAX} data
($7 \pm 1 \times 10^{20}$ $\rm cm^{-2}$, Haardt et al. 2001).

The unfolded spectrum of the {\it BeppoSAX} data is shown in
Figure \ref{fig1}.  The best fit model for BHSPEC with $i=67^{\circ}$ and
$\alpha=0.01$ which was used to generate the unfolded spectrum 
is also shown. The LMC X-3 spectrum
is very disk dominated and the PDS provides little constraint due to the low count
rate. Therefore, we fix the optical depth in the COMPTT model at $\tau=0.5$,
providing a flat (in $\nu F_{\nu}$) spectrum typical of thermal dominant state
observations.

We provide a comparison of the three spectral models
in Table \ref{tbl2}. With $D$ and $i$ fixed in the KERRBB and BHSPEC, each  model has 
the same number of free parameters: two for the soft/thermal component, one for the 
non-thermal component, and one for the intervening absorption column for a
total of four free parameters. DISKBB provides a
considerably poorer fit than either KERRBB or BHSPEC. The relativistically
broadened spectra are a much better representation of the soft thermal
emission than the narrower DISKBB.  

The quality of the DISKBB fit is sensitive to
the model of the non-thermal emission.  If we treat $\tau$ as a free parameter,
$\tau$ drops and
the fit improves slightly ($\chi^2_\nu=320/176$) but remains poor compared with KERRBB and BHSPEC.
The COMPTT spectrum steepens as $\tau$ decreases.  This extra flux in the `tail'
of the thermal component compensates for a decrease in $T_{\rm in}$ which allows
DISKBB to better approximate the low energy photons.  The fit further improves to
$\chi^2_\nu=264/176$ if we replace COMPTT with a power law. This provides a slightly
better fit than KERRBB but BHSPEC is still preferred.  The best fit model now
requires a steep power law component $\Gamma \sim 2.8$ which is consistent with
the best fit model of Haardt et al. (2001).  However, the power law flux now {\it
exceeds} the DISKBB flux at low energies.  This result is unphysical in a picture
where the soft X-ray emission provides the bulk of the seed photons for the
non-thermal component.  This is also likely the explanation for why the Haardt et
al. (2001) fits require a larger neutral hydrogen column.  This inconsistency
was also pointed out by Yao et al. (2005) who find a self-consistent fit with a
Comptonized multitemperature blackbody model.  In contrast, the inclusion of 
non-thermal emission has little effect on the $\chi^2$ values for KERRBB and
BHSPEC. Thus, the {\it BeppoSAX} data can be completely accounted
for by a bare accretion disk spectra as long as relativistic effects on the 
spectra are included.

BHSPEC provides a better fit ($\Delta \chi^2=-32$) to the data than KERRBB for an
inclination $i=67^{\circ}$, a source distance $D=52$ kpc, and $f=1.7$. Allowing
$f$ to vary from 1.5-1.9 does not improve the KERRBB quality of fit significantly.  
The prescription for relativistic effects in the two models are essentially
identical, so the differences in the spectral shapes are primarily due to the
different prescriptions for the disk surface emission.  The annuli spectra which
make up the BHSPEC model have imprints from
metal opacities and may differ from color corrected blackbodies by several
percent.  Additionally, annuli at different radii have local spectra which are best
approximated by different values of $f$, with $f$ usually being higher for the
hotter, inner annuli. KERRBB assumes one value of $f$ for the whole
disk.  Though these discrepancies are not at a level which is significantly
greater than the intrinsic uncertainties in the BHSPEC model, it is suggestive
that a model which includes atomic physics provides a better fit.

\begin{figure}
%\plotone{f3.eps}
\includegraphics[width=0.46\textwidth]{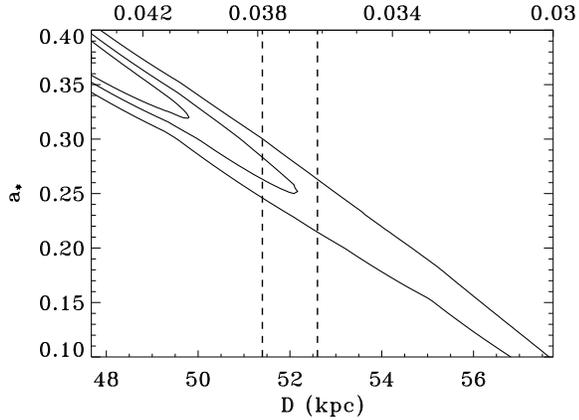}
\caption{The 66\%, 90\%, and 99\% confidence contours in the $a_{\ast}$--$D$ plane
of the best fit BHSPEC model ($\alpha=0.01$, $i=67^{\circ}$, $N$ free) to 
the {\it BeppoSAX} LMC X-3 data. Here we have let the normalization vary by 20\%
above and below its nominal value of $N=(10\,{\rm kpc}/52\,{\rm kpc})^2=0.0370$.
The model normalization $N$ is enumerated on the upper horizontal axis.  The 
vertical dashed lines mark uncertainty limits associated the distance estimate
in Table \ref{tbl1}
\label{fig3}}
\end{figure}

Despite these differences in the quality of fit, KERRBB and BHSPEC both give
values of $a_{\ast}\sim 0.3$ for $i=67^{\circ}$.  The best fit value for $a_{\ast}$ for BHSPEC
is a function of $\alpha$ with $\alpha=0.1$ giving a lower $a_{\ast}$ than
$\alpha=0.01$.  This is the case in all fits and is most simply understood by
examining how changes in the parameters either harden (increase the mean
photon energy) or soften (lower the mean photon energy) the spectra. As
$a_{\ast}$ increases, the inner radius of the disk decreases.  This results in a
larger fraction of the gravitational binding energy being released in a smaller
area of the disk surface.  The resulting increase in $\teff$ in
these annuli produces a spectrum with higher average photon energies. Therefore,
increasing $a_{\ast}$ hardens the spectra, even at fixed luminosity.

The sensitivity of the spectrum to $\alpha$ is more complex.  It 
is strongest at high $\mdot$
in radiation pressure dominated annuli where the surface density $\Sigma$ is low.
A larger $\alpha$ yields a lower $\Sigma$, making the disk less effectively optically thick.
For a range of  $\mdot$, the $\alpha=0.01$ annuli remain very effectively optically 
thick while the $\alpha=0.1$  annuli become less effectively optically thick and eventually
effectively optically thin as $\mdot$ increases.  For $\alpha=0.1$, the densities are
lower and the temperatures are higher. The photons cannot thermalize as well, causing 
the spectrum to harden significantly. At lower $\mdot$, both
models have sufficiently large $\Sigma$ so that spectral formation occurs nearer the disk
surface at approximately the same densities and temperatures.  The $\alpha=0.1$ models still 
tend to be slightly less dense in the spectral forming region and are therefore slightly harder,
but the differences are significantly smaller than at higher $\mdot$.
Therefore, the dependence of the best-fit $a_\ast$ on $\alpha$ results from an
increase in $\alpha$ from 0.01 to 0.1 hardening the spectrum so that $a_{\ast}$ must be
reduced to compensate. 

The best fit $a_{\ast}$ is also a function of $i$.  As can be seen in Table
\ref{tbl2}, making $i$ and $D$ free parameters does not significantly improve the
quality of fit in these cases.  However, it does greatly increase the
uncertainty in the best fit $a_{\ast}$.  This is illustrated in Figure
\ref{fig2} where we plot the 66\%, 90\%, and 99\% joint confidence
contours for $a_{\ast}$ and $i$.  The strong correlation exists because
lowering $i$ softens the spectrum so that $a_{\ast}$
must increase to compensate. This is partly because the line-of-sight projection of the 
azimuthal fluid velocity component decreases.  This reduces the blueshift and 
beaming of the emission from the
approaching side of the disk, and moves the `position' of the high energy tail
to lower energies. Also, the projected disk area increases which produces a
larger flux for the observer at infinity.  This needs to be compensated by a
decrease in $\mdot$, and therefore $\ell$.  This decrease in $\ell$
lowers $\teff$ and again softens the
spectrum. Thus, our ability to constrain $a_{\ast}$ is generally improved by
precise, reliable estimates for $i$.  The contours in Figure \ref{fig2}
suggest $a_{\ast} \simeq 0.55 \pm 0.2$ at 90\% confidence.  If the binary inclination
uncertainties are accurate and the inner accretion disk is aligned with the binary orbit,
these constraints imply $0.2 \lesssim a_{\ast} \lesssim 0.4$.

We also examine the variation of $a_{\ast}$ with the source distance $D$. The 
BHSPEC model normalization $N$ is defined so that $D=10/\sqrt{N}$ kpc. 
The confidence range for the
distance to LMC X-3 listed in Table \ref{tbl1} provides a tight constraint on $N$.
However, $N$ also depends on the absolute flux calibration of the detector so that 
any uncertainty in absolute flux translates into an effective uncertainty for $D$.  
Therefore, we consider fits where $N$ is free to vary by 20\% from its nominal values of 
$N=(10\,{\rm kpc}/52\,{\rm kpc})^2=0.0370$.  The 66\%, 90\%, and 99\% confidence 
contours in the $a_{\ast}$--$D$ plane are shown in Figure \ref{fig3}.  The best fit
model lies at the upper limit of the allowed range of $N$.  The best
fit $a_{\ast} \sim 0.38$ is slightly larger than the range ($a_{\ast} \sim 0.25 \pm 0.05$ )
consistent with the distance constraints which are plotted as vertical dashed lines in
the figure.

The strong anticorrelation between $a_{\ast}$ and $D$ seen in Figure \ref{fig3} exists
because an increase in $D$ leads to a decrease in the flux expected at the detector
(i.e. a lower $N$).  This must be accounted for by an increase in the luminosity ($\ell$)
which would shift the spectral peak to higher energies at fixed $a_{\ast}$. However, 
$a_{\ast}$ is a free parameter and it can be lowered so that the spectral peak remains 
fixed while $\ell$ increases.

\begin{deluxetable*}{lccccccc}
\tablecolumns{8}
\tablecaption{LMC X-3 {\it BeppoSAX} Fit Summary\label{tbl2}}
\tablewidth{0pt}
\tablehead{
\colhead{Model\tablenotemark{a}} &
\colhead{$\alpha$} &
\colhead{$i$} &
\colhead{$D$} &
\colhead{$a_{\ast}$} &
\colhead{$k T_{\rm in}$} &
\colhead{$N_H$} &
\colhead{$\chi^2_\nu$} \\
 &
 &
\colhead{(deg)} &
\colhead{(kpc)} &
 &
\colhead{(keV)} &
\colhead{$(10^{20}\,{\rm cm}^{-2})$} &
\\
}

\startdata
DISKBB &
\nodata &
\nodata &
\nodata &
\nodata &
$1.0139^{+0.0074}_{-0.0080}$ &
$4.18^{+0.21}_{-0.20}$ &
336/177 \\
KERRBB &
\nodata &
67 &
52 &
$0.3639^{+0.0013}_{-0.0013}$ &
\nodata &
$5.08^{+0.27}_{-0.25}$ &
275/177 \\
BHSPEC &
0.1 &
67 &
52 &
$0.141^{+0.021}_{-0.020}$ &
\nodata &
$5.65^{+0.27}_{-0.26}$ &
243/177 \\
BHSPEC &
0.01 &
67 &
52 &
$0.258^{+0.019}_{-0.019}$ &
\nodata &
$5.58^{+0.27}_{-0.26}$ &
246/177 \\
BHSPEC &
0.01 &
$53^{+13}_{-10}$ &
$51.4^{+1.2}_{-0.0}$ &
$0.54^{+0.11}_{-0.10}$ &
\nodata &
$5.92^{+0.31}_{-0.27}$ &
238/175 \\

\enddata

\tablenotetext{a}{The full Xspec model is WABS*(Model+COMPTT).}

\tablecomments{All uncertainties are 90\% confidence for one parameter.
Parameters reported without uncertainties were held fixed during the fit.}

\end{deluxetable*}

We also use these models to test for the possibility of magnetic torques on the
inner accretion disk. The energy release by a torque increases the fraction of
emission at small radii and increases the effective temperature of the annuli.  
This produces a hardening of the spectrum similar to an increase in $a_\ast$.  
At the time of publication, BHSPEC spectra have only been computed from disks with 
non-zero torques for $a_\ast=0$.  KERRBB can be used to examine torques at all
$a_{\ast}$, but the fits provide little constraint as $\Delta \eta/\eta$ varies over
the entire range of the model from zero to one at 90\% confidence when $i$ is a free
parameter.  As expected, increases in $\Delta \eta/\eta$ are offset by decreases 
in $a_\ast$.  An upper-limit on torque can be obtained by fitting BHSPEC at 
$a_\ast=0$.  The best fit $\Delta\eta/\eta=3 \pm 0.8$ with $\chi^2/\nu=245/175$ 
and $i=73^\circ \pm 1^\circ$. This value of $\chi^2$ is only slightly greater than
in the untorqued case, and the best fit inclination is consistent with the 
constraints on the binary inclination, so it is difficult to rule out the 
possibility of large torques from these data.

\subsubsection{{\it RXTE} Data} 

\begin{figure*}
%\plotone{f4.eps}
\includegraphics[width=0.94\textwidth]{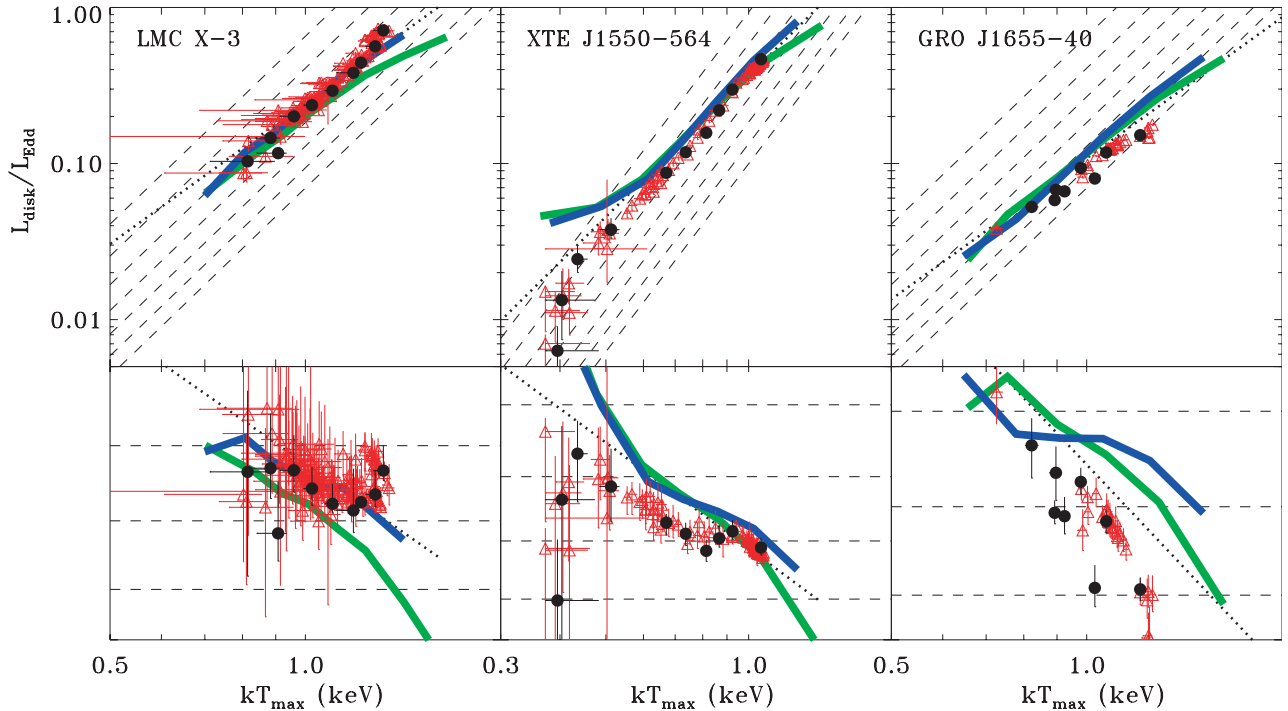}
\caption{The disk luminosity (top panels) as a function of maximum temperature 
for J1655, J1550,
and LMC X-3.  Each symbol represents a DISKBB fit to one {\it RXTE} data set. 
These data were presented in GD04 and the reader is referred there for a complete
discussion of their spectral analysis.  The black filled circles represent the
epochs which were used in this work and the red triangles represent the remaining
GD04 data sets.  The dashed curves are lines of constant color correction
corresponding to $f=1.6, 1.8, 2.0, 2.2, 2.4, 2.6$, and 2.8 for a Schwarzschild
black hole (see eq. 3 of GD04). The dotted lines represent curves with
$L_{\rm disk}/L_{\rm Edd} \propto T_{\rm max}^3$.  This $L-T$ relation follows
from simple, analytic estimates for the spectral hardening in effectively
optically thick disks when the effects of Comptonization are negligible. The
normalization is chosen arbitrarily to compare with the BHSPEC model curves. The
derivation of this relation and its relevance to the models is discussed
in the appendix. The green and blue curves show the evolution 
expected from the best fit BHSPEC models for $i$ fixed at the binary estimate
(see Table \ref{tbl1}) for $\alpha=0.1$ and 0.01,
respectively.  These curves were created by producing synthetic data sets with 
the BHSPEC model and replicating the spectral analysis performed by GD04. 
To more easily evaluate the spectral hardening relative to a fixed $f$,
we have plotted $L/T_{\rm max}^4$ (lower panels).  The units on the vertical
coordinates are arbitrary, but the lines of constant $f$ (now horizontal) are
retained for reference when comparing with the top panel.
\label{fig4}}
\end{figure*}

GD04 have already
selected a sample of disk dominated {\it RXTE} observations for several sources, including
LMC X-3. From these, we have selected a subset of 10 epochs which evenly cover
the range of disk luminosities inferred from the GD04 analysis.  The luminosities
of these epochs are plotted versus the maximum color temperature in the first panel
of Figure \ref{fig4}.  The black filled circles correspond to the data sets used
in our work and red triangles represent the other epochs in the GD04 sample.  
These plots were generated by taking DISKBB fit results and making corrections
for the temperature profile and relativistic effects (Zhang et al. 1997). This
plot only includes epochs in which the disk component is inferred to account for
greater than 85\% of the bolometric flux. A detailed explanation of the analysis
can be found in GD04. The values of $L_{\rm disk}/L_{\rm Edd}$ and $T_{\rm max}$
are evaluated using the estimates in Table \ref{tbl1} so there is some
uncertainty in collective position of these symbols on the plot.  However, the
placement of the points relative to each other is robust to these uncertainties
so that reproduction of the shapes of these $L-T$ relations provides an important test
for our disk models.  We repeat the same procedure for observations of J1550 
and J1655 and plot these in the center and right panels of Figure \ref{fig4},
respectively.

As stated in section \ref{intro}, the luminosity is roughly proportional to the
fourth power of the maximum temperature.  The dashed curves represent lines of
constant $f$ where $L \propto T_{\rm max}^4$. In the bottom panels of Figure
\ref{fig4} we have also plotted $L/T_{\rm max}^4$ in order to more easily evaluate
spectral hardening relative to this overall trend. Comparison of the data with these
curves shows some evidence for hardening with increasing $L$ for J1665 and J1550.  
LMC X-3 is roughly consistent with a constant $f$, but a close examination suggests 
there might be weak signs of hardening above $\sim 0.9$ keV and
softening at the highest temperatures.

We investigate this spectral evolution with $\mdot$ by fitting the models
directly to the data. We consider the same models as in section \ref{lmcsax}, but
we now fix the absorption column since it is not well constrained without the low
energy coverage.  For LMC X-3 we fix it at $N_H=5.5 \times 10^{20}$ $\rm cm^{-2}$
to be consistent with the {\it BeppoSAX} fits.
We initially fix $i$ and $D$, and fit only a single value of $a_\ast$ for all epochs.
We also fix $f=1.7$ for KERRBB and $\alpha=0.1$ or 0.01 for BHSPEC.  Only
$\mdot$ (or $\ell$) is allowed to vary for each epoch.
We also consider models with DISKBB, fitting a single 
normalization simultaneously to all data sets.  This is also consistent with assuming
a fixed color correction and constant effective area for each epoch. With these
choices, each model has the same number of free parameters.  There is a single parameter 
shared by all data sets ($a_{\ast}$ or DISKBB normalization), and two parameters
for each individual data set: one for the soft/thermal component ($\mdot$, $\ell$, or
$T_{\rm in}$) and a normalization for the non-thermal component.  For BHSPEC, we
also consider fits where $i$ and $D$ (or $N$) are free parameters.  Since only a
single value of either parameter is fit for all epochs, this provides at most
two additional parameters.

In oder to visualize the variation of the spectral shape with $\mdot$,
we plot in Figure \ref{fig4} the $L-T$ relations (solid curves) derived from
the best fit BHSPEC models at fixed $i$ for $\alpha$=0.1 and 0.01.  These curves are 
calculated by generating artificial spectra with our best fit models, and then fitting 
them using the same procedure that GD04 used for the real data.  The plot therefore
provides a comparison of fits with DISKBB, both to the data, and to artificial spectra
generated from the best-fit BHSPEC models.  It is {\it not} a direct comparison of BHSPEC
with the data.  This explains
why the best-fit BHSPEC curves do not go through the `data points' in the J1550 and J1655 
plots.  The two types of fits find a different partition of the spectra between the 
soft/thermal and hard/non-thermal components, with additional flux accounted for
by the non-thermal emission in the BHSPEC fits. The implications of this are discussed
further in section \ref{hard}. For comparison, we also show curves with 
$L_{\rm disk}/L_{\rm Edd} \propto T_{\rm max}^3$ (dotted lines).  These curves
represent simple, analytic estimates for the spectral hardening in effectively
optically thick disks when the effects of Comptonization are negligible.  The
derivation of this relation in presented in the Appendix.

Based on the discussion above, it is reasonable to expect that
the difference in quality of fit between models with fixed $f$ (KERRBB and DISKBB)
and BHSPEC might be dominated by the differing predictions for the variation of spectral
hardening with changing $\ell$.
The lack of hardening at the highest $\ell$ seen in Figure \ref{fig4} for LMC X-3 suggests 
that models with constant $f$ or BHSPEC with a low value of
$\alpha$ would provide the best representations of the data. These
predictions are borne out in simultaneous fits to the LMC X-3 data, which are
summarized in Table \ref{tbl3}. The unfolded spectra are plotted in
Figure \ref{fig5} with the best fit BHSPEC model for $\alpha=0.01$ and
$i=67^{\circ}$.  DISKBB, which is representative of a disk with a fixed emitting
area and constant $f$, provides the best fit. A comparable $\chi^2$ is provided by
KERRBB with $i=67^{\circ}$ and $f=1.7$.  The fit with BHSPEC for $\alpha=0.01$
gives an acceptable $\chi^2$, but provides a poorer representation than the fixed 
$f$ models. The $\alpha=0.1$ model does not provide an acceptable fit.
For BHSPEC, the largest contributions to $\chi^2$ come from the highest luminosity
epoch.  As seen in Figure \ref{fig4}, the BHSPEC models seem to harden too rapidly to
accommodate all epochs simultaneously. BHSPEC hardens more rapidly with 
increasing $\ell$ for $\alpha=0.1$ than 0.01, leading to the significantly
poorer fit.  In order to make the inner disk annuli more
effectively thick, we increased the metal abundances to three times the solar
value.  However, this had limited impact on the spectral shape and did not
improve the quality of fit appreciably.

The best fit values of $a_{\ast}$ and their 90\% confidence intervals are
summarized in Table \ref{tbl3}.  They are systematically lower ($a_{\ast} \lesssim
0.1$) than the values inferred from the {\it BeppoSAX} fits ($a_{\ast}\sim 0.3$).  
Most of this discrepancy can be accounted for by cross calibration differences
between the two observatories.  Cross-calibration\footnotemark \,  campaigns on 3C 273
show that the {\it RXTE} PCA flux is about 20\% higher than the {\it BeppoSAX} data.
At fixed normalization, this requires a $20\%$ increase in $\ell$ which leads
to an $\sim 5\%$ increase in $\teff$.  For a given $i$, this change must be offset by 
a decrease in $a_\ast$ which leads to a systematically lower value for the {\it RXTE}
fits relative to {\it BeppoSAX} fits.

\footnotetext{http://heasarc.gsfc.nasa.gov/docs/asca/calibration/3c273\_results.html}

\subsection{XTE J1550-564}
\label{1550}

As seen in the middle panels of Figure \ref{fig4}, J1550 varies over an order of
magnitude in $\ell$ and we have chosen 10 observations which sample this range.  We
fit J1550 (and J1655) with the same model which we applied to LMC X-3, but we
found statistically significant residuals consistent with reflection features.
(These were apparently unnecessary in LMC X-3 because the spectra are so strongly
disk dominated.)
To account for these residuals we add a GAUSSIAN and apply a SMEDGE (Ebisawa et
al. 1991) to the COMPTT component to approximate reprocessing at the disk
surface. We fix the widths of these components to 0.5 keV and 7 keV,
respectively.  This adds two free parameters from each new component.  We also
let the optical depth vary in the COMPTT component for a total of
five additional parameters in each epoch. The fit results with these additional
components are summarized in Table \ref{tbl3}.

As with LMC X-3, all of the models provide an acceptable fit, though there is
still considerable variation in the $\chi^2$ values.  There is a statistically
significant preference for the DISKBB model over KERRBB.  Since both models have
a constant $f$, the difference in the quality of fit must be due to the overall
spectral shape and not simply its evolution with $\mdot$.  A comparison of
the best fit spectral shapes shows that relativistic KERRBB models are harder
than those of DISKBB over the {\it RXTE} band, which seems to be primarily the
result of relativistic
broadening.

The BHSPEC model with $\alpha=0.01$ provides a better fit than $\alpha=0.1$,
consistent with the prediction of the $L-T$ comparison in Figure \ref{fig4}.  
However, the $\alpha=0.01$ model still provides a poorer fit than DISKBB.  In
contrast to KERRBB, the best fit spectra fall off more steeply with increasing photon
energy than the DISKBB spectra.  This behavior seems to be primarily due to
absorption features (primarily Fe K) in the tail of spectrum. When we let $i$
and $D$ float, the BHSPEC fit improves and the $\chi^2$ values are now slightly
better than DISKBB. The best fit inclination is lower ($i=42^{+3}_{-13}$ deg)
and the spin is higher ($a_{\ast}=0.72^{+0.15}_{-0.01}$),
producing slightly harder spectra with less pronounced absorption features than
in the $i=72^{\circ}$ case.
For $i=72^{\circ}$, the best fit $a_{\ast}$ is relatively low ($\lesssim 0.1$)
for both KERRBB and BHSPEC, so allowing $i$ to vary changes $a_{\ast}$ considerably.

Apparent motion $\gtrsim 2 c$ has been claimed to be observed in the radio
emission from this source (Hannikainen et al. 2001) suggesting $i \lesssim
50^{\circ}$ for ballistic motion.  This implies a misalignment of at least
$20^{\circ}$ if the jet is aligned with the black hole angular momentum vector.
The best fit inclination $i=42^{+3}_{-13}$ deg is consistent with this upper
limit and may be compatible with an inner disk aligned with the black hole via
the Bardeen-Petterson effect.

\subsection{GRO J1655-40}
\label{1655}

Of the three sources considered in this work, J1655 displays the strongest
evidence of hardening in Figure \ref{fig4}, suggesting that the models with fixed
$f$ will provide poorer fits than BHSPEC.  In fact, KERRBB does not provide an
adequate fit to the data, and a luminosity dependent trend can be seen in the
residuals. However, DISKBB can still provide an adequate fit and is even
preferred to BHSPEC for $i=70^{\circ}$.  As in J1550, the best fit KERRBB spectra
are harder than those of DISKBB. These results again suggest that the overall
differences in spectral shape (as opposed to the variation with $\mdot$)
provide the dominant effect on the quality of fit.  Unlike LMC X-3 and J1550, the
BHSPEC model with $\alpha=0.1$ provides a better fit than $\alpha=0.01$.
Comparison with the bottom-right panel of Figure \ref{fig4} suggests that the soft, thermal
component in J1655 is hardening more strongly with increasing luminosity than can
be easily accounted for with the $\alpha=0.01$ model.

At fixed inclination, BHSPEC with $\alpha=0.1$ provides the only relativistic fit
which is marginally acceptable. The best fit $a_{\ast}=0.62 \pm 0.01$ in this
case suggests a moderate spin which is reasonably consistent with other
investigations ($a_{\ast}\sim 0.7-0.9$, Gierlinski et al. 2001; $a_{\ast}\sim
0.65-0.75$, Shafee et al. 2006).  KERRBB also yields a similar spin 
($a_\ast=0.6015^{+0.0013}_{-0.0023}$) for $f=1.7$.  The small discrepancy with
the Shafee et al. (2006) is likely due to our choice of a single $f$ and
to differences in our modeling of the non-thermal emission.

Allowing $i$ to be a free parameter significantly reduces
$\chi^2$ for both values of $\alpha$, but $\alpha=0.01$ now provides a slightly
better fit. As with LMC X-3 and J1550, the spin is very sensitive to the inclination.
For both values of $\alpha$ the best fit $a_{\ast}=0$, the lower limit of the model.
(We have not yet extended BHSPEC to retrograde spins.)  
We find $\chi^2_\nu=248/302$ and $i=83.8^{\circ} \pm 0.6^{\circ}$ for
$\alpha=0.1$, and $\chi^2_\nu=234/302$ and $i=85.6^{\circ} \pm 0.4^{\circ}$
for $\alpha=0.01$.

Radio observations of J1655 have inferred a jet inclination of $85^{\circ} \pm 2^{\circ}$
to the line of sight (Hjellming \& Rupen 1995). Assuming the jets are aligned
with the angular momentum axis of the black hole, this would imply an inner disk
inclination (to the plane of the sky) of $i=85^{\circ}$, consistent with the best
fit $i$ above.  However, disk alignment only occurs for black holes with
non-zero angular momentum and the transition radius is expected to increase with
increasing spin (Bardeen \& Petterson, 1975).  Therefore, because these fits find
$a_\ast=0$, they do not provide a self-consistent picture for a misaligned disk
scenario.

\begin{deluxetable*}{lccccccc}
\tablecolumns{8}
\tablecaption{{\it RXTE} Fit Summary \label{tbl3}}
\tablewidth{0pt}
\tablehead{
 &
 &
\multicolumn{2}{c}{LMC X-3} &
\multicolumn{2}{c}{XTE J1550-564} &
\multicolumn{2}{c}{GRO J1655-40} \\
\colhead{Model\tablenotemark{a}} &
\colhead{$\alpha$} &
\colhead{$a_{\ast}$} &
\colhead{$\chi^2_\nu$} &
\colhead{$a_{\ast}$} &
\colhead{$\chi^2_\nu$} &
\colhead{$a_{\ast}$} &
\colhead{$\chi^2_\nu$} \\
}

\startdata
DISKBB\tablenotemark{b} &
\nodata &
\nodata &
296/431 &
\nodata &
230/355 &
\nodata &
284/304 \\
KERRBB\tablenotemark{c} &
\nodata &
$0.119^{+0.013}_{-0.013}$ &
296/431 &
$0.097^{+0.005}_{-0.065}$ &
301/355 &
$0.6015^{+0.0013}_{-0.0023}$ &
439/304 \\
BHSPEC &
0.1 &
$0$ &
1190/431 &
$0^{+0.0055}_{-0}$ &
324/355 &
$0.617^{+0.013}_{-0.006}$ &
330/304 \\
BHSPEC &
0.01 &
$0^{+0.006}_{-0}$ &
359/431 &
$0.115^{+0.030}_{-0.011}$ &
256/355 &
$0.639^{+0.012}_{-0.006}$ &
369/304 \\
BHSPEC\tablenotemark{d} &
0.01 &
$0.728^{+0.036}_{-0.018}$ &
307/429 &
$0.72^{+0.15}_{-0.01}$ &
221/353 &
$0.00^{+0.021}_{-0}$ &
234/302 \\
\enddata

\tablenotetext{a}{The full Xspec model is WABS*(Model+COMPTT) for LMC X-3, and
WABS*(Model+GAUSSIAN+SMEDGE*COMPTT) for J1550 and J1655.}
\tablenotetext{b}{A single normalization is fit for all data sets.}
\tablenotetext{c}{The hardening factor is fixed at $f=1.7$. Fit parameters were 
selected so that limb darkening was included but self-irradiation was not.}
\tablenotetext{d}{Both $D$ and $i$ are free parameters with the value of $i$ 
allowed to vary over the full range but $D$ constrained to lie within the 
confidence limits reported in Table \ref{tbl1}. The best fit values are $
i=43.4^{+8.5}_{-4.0}$ deg,
$D=51.4^{+0.61}_{-0}$ kpc for LMC X-3; $i=42^{+3}_{-13}$ deg
$D=6.3510 \pm 0.0080$ kpc for J1550; and $i=85.60^{+0.14}_{-0.38}$ deg, 
$D=3.4000^{+0}_{-0.0059}$ kpc for J1655.}  

\tablecomments{All uncertainties are 90\% confidence for one parameter.
Parameters reported without uncertainties were held fixed during the fit.
Unless otherwise noted, the values of $i$, $D$, and $M$ were fixed at the 
estimates given in Table \ref{tbl1}.}

\end{deluxetable*}

\section{Discussion}
\label{discus}

One of the principle aims of this work was to test the applicability of the
relativistic $\alpha$-disk model in BHBs. The spectral fitting discussed in
section \ref{specmod} presents mixed results.  The significant improvements in
$\chi^2$ relative to DISKBB resulting from the BHSPEC and KERRBB fits to the {\it
BeppoSAX} data provide a strong case for relativistic broadening.  DISKBB alone
is too narrow to adequately approximate the soft, thermal emission from LMC X-3.  
The additional quality of fit improvement for BHSPEC relative to KERRBB might
also be evidence for modified blackbody and smeared absorption features in the
spectrum.

In light of these results, it is surprising that DISKBB with a fixed
normalization seems to provide a better fit to the {\it RXTE} data than KERRBB
or BHSPEC in all three sources when we fix $i$ at the binary inclination. Since
it is preferred to both relativistic models, the difference cannot simply be due
to the differences in the degree of spectral hardening as luminosity changes. This could be
taken as evidence against relativistic broadening, but the innermost radii
implied by the DISKBB fits are consistent with coming from near the black hole.  
A comparison of the best fit spectral shapes for all three models shows
differences at the $\lesssim 10\%$ level.  The KERRBB spectral shapes tend to be
broader (harder in the 3-20 keV band) than both DISKBB and BHSPEC. The
differences between BHSPEC and KERRBB seem to be mostly due to broad absorption
features which cause the BHSPEC model to fall off more strongly with increasing
energy in the tail of the spectrum.  Therefore, it is conceivable that DISKBB
spectrum could be mimicking similar, but slightly weaker, features in data,
though it is surprising that it does this consistently and effectively in all
three sources.

Alternatively, it may simply be that our estimates for $D$, $M$, or $i$ are in
error.  When we allow $i$ to be a free parameter, BHSPEC
provides a better fit than DISKBB in both J1550 and J1655.  In 
both cases, observations of the radio jets
suggest the angular momentum of the black hole is misaligned with that of the
binary. It is suggestive that in both cases we find values for $i$ consistent
with the constraints implied by the jets, rather than the binary inclination.
In the case of LMC X-3, where a jet has not been observed, the best fit $i$
is more nearly face on than the measured binary inclination.  It is consistent
with the binary inclination at 90\% confidence for the {\it BeppoSAX} data, but
not for the {\it RXTE} spectral fits.

A third possibility is that the non-thermal emission and Compton reflection 
components are not being correctly accounted for by our prescription.  If
this is a problem, it should be minimized by looking at LMC X-3, which has
the most disk dominated spectra of the three sources and relatively little
evidence for reflected emission.
For LMC X-3 the BHSPEC residuals are clearly dominated by the most luminous epoch
for which BHSPEC predicts too much hardening with increasing luminosity for
either value of $\alpha$. If we ignore the most luminous epoch $\chi^2_\nu$
improves to 257/388 for BHSPEC with $\alpha=0.01$ and $i=67^{\circ}$, but only
improves slightly to 256/388 for DISKBB, providing comparable fits. Allowing $i$ 
to be a free parameter allows the BHSPEC fit to improve even further for slightly
more face on values.  If $i$ remains fixed at $67^{\circ}$, the fit with BHSPEC also
improves by allowing the model normalization to vary within 20\% of the nominal
value.  This is a larger range of normalization than that associated with
distance uncertainty and accounts for possible errors in the absolute flux
calibration. Thus for LMC X-3, the shape of the spectra seem to be
best represented by BHSPEC, but the evolution of the spectral hardening with
$\ell$ at the highest luminosities is not consistent with the predictions of a
simple $\alpha$-disk model.

\subsection{Spectral Hardening and the Stress Prescription}

The three $L-T$ relations presented in Figure \ref{fig4} show that differences
exist in the spectral evolution with disk luminosity from source to source.  LMC
X-3 has the most scatter and is reasonably consistent with a constant $f$,  
though the lower-left panel of Figure \ref{fig4} shows evidence of a weak
hardening for $0.1 \lesssim \ell \lesssim 0.3$ and softening for $\ell \gtrsim 0.3$.
J1550 is consistent with weak hardening and J1655 seems to show more significant
hardening.  A comparison of the KERRBB and BHSPEC fit results seems to agree with
these descriptions.  At fixed $i$, KERRBB with $f=1.7$ is preferred for LMC X-3, but
BHSPEC provides a better fit in both J1550 and J1655.  Fitting a single $f$ for
all epochs with KERRBB does not alter this result.  A comparison of BHSPEC fits
with $\alpha=0.1$ and 0.01 is also generally consistent. With $i$ fixed at the
estimate for the binary inclination, $\alpha=0.01$ is preferred for LMC X-3 and
J1550, but $\alpha=0.1$ provides a better fit for J1655.  If $i$ is a free
parameter, $\alpha=0.01$ provides a better fit for all three sources, though the
improvement is small for J1655.

As discussed in section \ref{lmcsax} and the Appendix, the sensitivity of the 
spectra to $\alpha$ comes about  primarily because $\alpha$ determines $\Sigma$.
For sufficiently  large $\Sigma$, the spectral shape depends only weakly on $\alpha$.
However, if $\Sigma$ drops sufficiently, the disks begin to become effectively optically 
thin at small radii. Once the hottest annuli are effectively optically thin, they become 
increasingly isothermal or inverted in their temperature profiles as inverse Compton
scattering in the now hotter surface layers increasingly dominates the cooling.
This is much less efficient than thermal cooling and the spectra harden rapidly as
temperatures rise with increasing $\teff$ (Davis \& Hubeny 2006).  Such 
effects are responsible for the hardening in the $\alpha=0.1$ models at high 
$\ell$ in Figure \ref{fig4}.  Our results
suggest that these multi-epoch fits are sensitive to these effects and may
even be able to differentiate between the  $\alpha$-disk prescription
and more general models of angular momentum transport in these disks.

In the context of the $\alpha$ prescription, our fit results seem to rule out
fixed values of $\alpha \ge 0.1$ for LMC X-3 and possibly J1550.
At first sight, this appears inconsistent with the $\alpha \ge 0.1$ values
inferred for the outburst phases of dwarf novae (Lasota 2001) and soft
X-ray transients (e.g. Dubus et al. 2001).  However, it is important to note
that the disk instabilities that drive the outburst time scales are
associated with regions of the disk where gas pressure dominates radiation
pressure.  In contrast, the X-ray spectra are dominated by the innermost
regions of the disk where radiation pressure can be important at high
$\mdot$.  There is no reason to believe that either $\alpha$ or the
stress prescription should be the same at all radii in the disk.

It may be that the classical stress prescription of equation (\ref{taurphi})
is not valid in radiation pressure dominated disks.  Alternative stress
prescriptions have long been considered, partly
because they can produce thermally and viscously stable disks (e.g. Piran
1978).
In addition, there have been proposals that magnetohydrodynamical
turbulence may produce stresses that are limited to values that are related
in some way to the gas pressure (e.g. Sakimoto \& Coroniti 1981, Merloni 2003).

It is noteworthy that
LMC X-3 reaches the highest Eddington ratios among the sources fit here, and that
the spectrum even appears to {\it soften} slightly at these highest luminosities.
A substantial reduction in stress at $\sim 0.6-0.7$ might explain this, perhaps linked
to the onset of a disk instability, which appears to be present in the only other
source to consistently exceed such luminosities, GRS 1915+105.  Fits with BHSPEC
to GRS 1915+105 show that this source has `stable' disk spectra (i.e. constant
for longer than 16 sec) from $\ell=0.5-0.6$ and from $\ell=1-2$ which are consistent with
the expected hardening with luminosity (Middleton et al. 2006).
However, there are {\it no} stable disk spectra from this source in the range
$\ell \sim 0.7-0.9$, exactly where the LMC X-3 spectra show slightly different
properties than expected.

Other effects might be important at the high luminosities.
The disk models used in BHSPEC are actually somewhat inconsistent for
$\ell \gtrsim0.3$, as the innermost annuli then have
$H/R\gtrsim0.1$.  Radial transport of accretion power is increasingly
important in this regime.  In addition, magnetic torques across the ISCO
may be more important (Afshordi \& Paczy\'nski 2003).  However,
both of these effects would tend to make the inner annuli hotter and/or
more effectively optically thin. We would expect this to {\it increase}
the hardening, in contrast to what is observed.  As the Eddington ratio
increases, increased inhomogeneities in the magnetorotational turbulence 
(Turner et al. 2002) and photon bubbles (Begelman 2001) may mitigate this by producing a
softer spectrum and a geometrically thinner disk than would be expected in a homogeneous model.

\begin{figure}
%\plotone{f5.eps}
\includegraphics[width=0.46\textwidth]{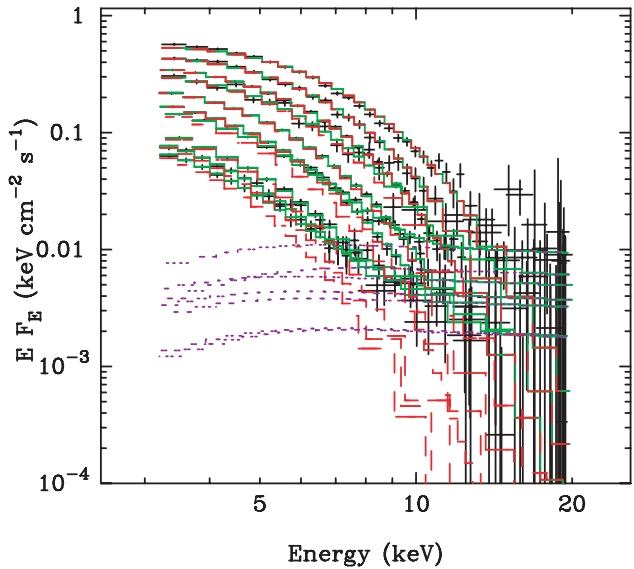}
\caption{The unfolded spectra for the {\it RXTE}
observations of LMC X-3 using the best fit BHSPEC models for $i=67^{\circ}$, 
$D=52$ kpc,and $\alpha=0.01$.  The total model component (green, solid curve),
BHSPEC (red, long-dashed curve) and COMPTT (violet, short-dashed curve) are
plotted.
\label{fig5}}
\end{figure}

\subsection{Estimates for Black Hole Spins}

The total mass accreted over the lifetime of these sources expected to be
small (King \& Kolb 1999; Shafee et al. 2006). As a result, only a
small increase in the angular momentum of the black hole is expected, and so our 
spin measurements are likely probing the natal spin 
distribution of the binaries. Therefore, these low to moderate spin estimates
($a_\ast \lesssim 0.8$) place constraints on black hole formation scenarios.  They
may also constrain spin-dependent models for jet production (Middleton et al. 2006)
since both J1550 and J1655 are microquasars. We characterize these spins
as `moderate' because even for $a_{\ast}\sim 0.8$,
the proximity of the ISCO to the event horizon
and the resulting radiative efficiency ($R\sim 3 R_g$, $\eta\simeq 0.12$) are 
substantially less extreme than in the maximally spinning spacetime 
($a_{\ast}\sim 0.998$, $R\sim 1.24 R_g$, $\eta\simeq 0.32$).  These moderate spins
are in contrast to estimates for spin at or near $a_\ast=0.998$ that have been
inferred from other methods, including spectral fits to the Fe K$\alpha$ line and 
some interpretations of the high frequency quasi-periodic oscillations (QPOs).

Broad Fe K$\alpha$ line emission has been seen in {\it ASCA} observations of
J1550 and J1655 (Miller et al. 2004).  The emission was modeled with a
relativistic disk line profile calculated in a maximally spinning Kerr spacetime
(LAOR in Xspec, Laor 1991), but with inner radius allowed to vary.  The best fit
inner radius for J1655 was small with $R_{\rm in} \lesssim 2 R_g$, suggesting
$a_\ast > 0.9$ and possibly near maximal ($a_\ast \sim 0.998$).  J1550 is less 
well constrained with $R_{\rm in} \lesssim 4-6 R_g$ depending on the model, though
it may also be consistent with near maximal spin.
The strongest constraints on $a_\ast$
come from the presence of a broad, asymmetric red wing which extends down to
$\sim$4 keV in the best-fit models.  Therefore, a principle source of
uncertainty for applying this method is modeling the underlying continuum to
accurately gauge the shape and extent of the line wing.  We refer the reader to
Done \& Gierlinski, (2005) for a more detailed discussion of uncertainties associated
with this method.

The reproducibility of the frequencies of the pair of high frequency QPOs from one
observation to the next suggests they might also provide a direct probe of the
black hole spacetime.  As a result, prospective models of QPOs often provide
constraints on $a_\ast$. For example, if the lower frequency member of the pair
in J1550 and J1655 is identified as an
axisymmetric radial epicyclic oscillation, then black hole spins of $a_\ast >
0.9$ are required (Rezzolla et al. 2003, T\"or\"ok 2005).  The reason is simple:
the radial epicyclic frequency has a maximum at some radius, and that maximum is
below the observed QPO frequency for the observed black hole masses unless the
spin is high.  The same conclusion holds in diskoseismology models if the lower
frequency member is identified with a low order axisymmetric ``g-mode'', because
it has a frequency less than the radial epicyclic frequency within the mode
trapping region (e.g. Wagoner et al. 2001). However, there are many possible
oscillation modes in accretion disks, and other mode identifications can be made
which are more consistent with moderate spins (e.g. Blaes et al 2006a).

In principle, the high signal-to-noise {\it RXTE} data allow us to place
very tight constraints on the spins of BHBs (see e.g. Table 3).  However, the
relatively small uncertainties in Table 3 do not account for uncertainties in
$M$, $i$, and $D$.  There are degeneracies among the fitting parameters in the way
that they affect the spectrum, e.g. the correlation
between $i$ and $a_\ast$ seen in Figure \ref{fig2}. Therefore, when $i$ and $D$ are
free parameters, $a_\ast$ can 
change significantly.  For the {\it RXTE} data, the uncertainty ranges for 
$a_\ast$ increase when $i$ and $D$ are free, but still remain relatively small.
This also leads to rather small formal uncertainties for $i$ and $D$, though 
$D$ is at the limit of allowed range for LMC X-3 and J1655.
So these data are capable of constraining all the parameters simultaneously, 
because changes of only a few percent in the spectral shape in the high energy
tail of the spectrum lead to substantial 
changes in $\chi^2$.  However, we caution that the models themselves are uncertain
at the few percent level due to our interpolation method alone (Davis \& Hubeny 2006).
Therefore, this method of spin estimation can
only be used with good precision when reliable and  precise constraints on 
$M$, $i$, and $D$ are available.

All of these methods of spin estimation have their weaknesses, as all rely to 
varying degrees on uncertain physical assumptions. An important example relevant
for both Fe K$\alpha$ line and our continuum spectral fits is the assumption
that the disk emission extends to the ISCO, and effectively ceases interior to
this radius. In principle, the disk (or its emission) could be truncated at larger
radius and our spins would be underestimates.  Alternatively, significant
emission might be generated or reprocessed inside the ISCO (Krolik \& Hawley 2002), 
making the interpretation of both estimation methods more difficult.

Several uncertainties also remain in other physical assumptions which underly
the BHSPEC model, because we still lack a complete
understanding of the magnetohydrodynamical structure of the accretion flows.
There are several modifications which likely lead to a hardening of the spectra,
including
increased dissipation near the disk surface (Davis et al. 2005), an increase
in the density scale height due to magnetic pressure support (Blaes et al. 2006b), 
surface irradiation by the non-thermal emission, and torques on the
inner edge of the disk.  Other processes might lead to a softening of the spectra.
Opacity due to bound-bound transitions of metal ions, which is not included in
BHSPEC should increase
the ratio of absorption to electron scattering opacity, pushing the
spectrum closer to blackbody. Inhomogeneities, such as those caused by 
compressible magnetohydrodynamical turbulence (Turner et al. 2002) or the photon bubble
instability (Turner et al. 2005), may also soften the spectrum.  They may increase the
effective ratio of 
absorption to scattering opacity because photon-matter interactions are
dominated by the densest regions (Davis et al. 2004), or through a reduction 
in the density scale height (see e.g. Begelman 2001) leading to an increase in
the average density of the disk interior.  Given all these possibilities, it is 
difficult to say with certainty what net effect modifying our assumptions or
including these additional processes would have on the disk spectra. If the effects
which harden the spectra are more important, BHSPEC 
will underestimate the spectral hardening and would then require 
higher spins to fit the data, making our spin measurements overestimates
and vice-versa.

Given these uncertainties, it is conceivable that one could reconcile the
spectral constraints with $a_\ast \sim 0.9$ if the BHSPEC model overestimates the
actual spectral hardening.  Spins this high would bring our estimates in
line with the published uncertainties for the Fe K$\alpha$ estimates and some
of the QPO-based measurements. Reconciling our spectral models
with the extreme spin ($a_\ast \sim 0.998$) is more difficult.  We have
attempted to quantify this in the case of J1655 by fitting the data with KERRBB for
$a_\ast=0.998$, but with $f$ as a free parameter.  For $i=70^{\circ}$ or $85^{\circ}$, 
an adequate fit can be obtained only for $f \sim 1$ which corresponds to blackbody 
radiation.  Electron scattering opacity dominates at the relevant temperatures so 
nearly blackbody emission is highly unlikely. One could obtain agreement
with more reasonable color corrections ($f \sim 1.5$)
by allowing $i$ to be a free parameter.  However, this requires $i\lesssim 40^{\circ}$,
in disagreement with the inclinations from both the binary and jet observations.

\subsection{Uncertainties due to Hard X-ray Emission}
\label{hard}

A potential difficulty for deriving constraints on $\alpha$ or $a_\ast$ by
this method is the
need to account for the non-thermal emission.  This is particularly true with {\it RXTE}
data when only the high energy tail of the spectrum is typically observed. 
Even though we infer the  models' bolometric
fluxes to be dominated by the softer thermal component, the
non-thermal component accounts for a substantial fraction of the 3-20 keV
emission in many of the cases.  The decomposition of thermal and non-thermal
emission is clearly dependent on the choice of model for the thermal emission.  
It can be seen in Figure \ref{fig4} that the $T_{\rm max}$ values derived from the
best-fit BHSPEC models for J1550 and J1655 are softer (lower $T_{\rm max}$) at
fixed $L_{\rm disk}/L_{\rm Edd}$ than those derived by GD04 who fit DISKBB directly 
to the data.  This means
that emission which was being accounted for by the DISKBB model is partly being
accounted for by the non-thermal emission in the BHSPEC fits.  The fits also
depends on the choice of model for the non-thermal component.  The COMPTT model
assumes a single Wien spectrum for the soft photon input.  A model with a
multitemperature disk spectrum for the seed photon input (THCOMP, Zdziarski et al. 1996)
provides more low energy photons for a given temperature.  Therefore, a THCOMP
spectrum which matches COMPTT at higher photon energies will tend to have more
flux at lower energies.  We find that the quality of fit can change significantly
(e.g. DISKBB is no longer preferred to BHSPEC for J1550 at fixed $i$) if we
replace COMPTT with THCOMP, though the best-fit spins and preferred $\alpha$ values
seem more robust.  The LMC X-3 observations are much more disk dominated 
than those of J1550 and J1655, and are much less sensitive
to the choice of model for the non-thermal emission.

The effects of the non-thermal emission can be minimized if we
observe these sources with detectors sensitive to lower photon energies, as  most
of the spectra peak below the lower limit of the {\it RXTE}
band. Thus, the {\it RXTE} fits can be sensitive to changes of only a few percent
in both the non-thermal emission model and the shape of the high energy tail of 
the BHSPEC models. The shape of the high energy tail of the BHSPEC spectra is uncertain at
the several percent level due to our interpolation scheme (Davis \& Hubeny 2006).
Line-of-sight absorption at softer-photon energies will
put practical limits on such modeling, and makes sources located
out of the Galactic plane (such as LMC X-3) particularly well suited for this
type of study.

\section{Conclusions}
\label{conc}

We analyze disk-dominated spectra of three black hole binaries: LMC X-3,
XTE J1550-564, and GRO J1655-40, fitting them with a simple multitemperature
blackbody model (DISKBB), as well as sophisticated relativistic disk models
(KERRBB and BHSPEC).  For LMC X-3, this includes {\it BeppoSAX} data, which
cover the 0.1-10 keV energy band in which the majority of the bolometric
flux of the disk is emitted.  In this case we find a statistically significant
preference for the relativistic models over DISKBB.  At lower significance,
we also find a preference for BHSPEC over KERRBB which may suggest the
spectra are sensitive to atomic and radiative transfer physics which are calculated
explicitly in BHSPEC.

We also examine {\it RXTE} spectra for each of the three BHBs using simultaneous,
multi-epoch fits to each source.  When we fix the relativistic model at the 
independent estimates for the binary inclination in Table \ref{tbl1}, we find DISKBB
is preferred over both relativistic models for both J1550 and J1655, though $\chi^2_\nu$ is
still acceptable in most of the relativistic model fits. If we allow the inclination 
to be a free parameter, BHSPEC is the best-fit model for both sources.  The best fit 
inclinations are both consistent with constraints inferred from radio observations
of the jets in these sources which might be accounted for by a misalignment of 
the black hole spin with the binary orbital angular momentum.  BHSPEC is also
the best fit model for LMC X-3, if we ignore the highest luminosity epoch
where additional physics appears to be important.  The best-fit inclination is marginally
consistent with the constraints on the binary inclination in this source.

Using the binary inclination estimates in Table \ref{tbl1}, we are able to derive
precise estimates for the black hole spin. However, the inferred values of the spin 
are functions of inclination, and the spin changes if the inclination of the X-ray
emitting, inner disk annuli differs significantly from independent estimates.
The best fit spin is also sensitive to the
source distance and absolute flux calibration through the model normalization.
Accurate and precise estimates for all these parameters as well as the black
hole mass are therefore a prerequisite for accurate and precise spin estimates.
We find relatively moderate spins ($a_\ast \lesssim 0.8$), even when inclination
is a free parameter in our fits. For J1655, our maximum spin estimate is only slightly 
lower than the limits ($a_{\ast} \gtrsim 0.9$ at 90\% confidence) implied by fits to 
Fe K$\alpha$ lines 
(Miller et al. 2004) and certain models of high frequency QPOs (
e.g. Wagoner et al. 2001; Rezzolla et al. 2003; T\"or\"ok 2005).  The spin of J1550
is more weakly constrained by the Fe K$\alpha$ fits (Miller et al. 2004), and is 
consistent with our estimates. However, for both sources the best fit Fe K$\alpha$
models are also consistent with nearly `maximal' spins ($a_{\ast}\sim 0.998$) which 
would be difficult to reconcile if the BHSPEC models provide an accurate approximation
to the spectra of the accretion flows in these sources.

We also find that our fits are sensitive to the assumed form of the angular momentum 
transport through its effects on the disk surface density. We consider an $\alpha$ stress 
prescription with $\alpha=0.1$ and 0.01, finding approximate qualitative
agreement between the $L-T$ diagrams and the model predictions (see figure
\ref{fig4}).  We find that all three BHBs are consistent with a single value
of $\alpha$, preferring $\alpha=0.01$ to $\alpha=0.1$
(though only weakly in the case of J1655).  These results are in contrast to the 
standard disk instability model of soft X-ray transients (see e.g. Dubus et al. 2001)
which requires $\alpha \gtrsim 0.1$ in the outer disk.
If we include the most luminous epoch in our fits to the 
{\it RXTE} data of LMC X-3, models with constant $f$ provide a better
fit than BHSPEC with either $\alpha$. BHSPEC fails to account for the most
luminous epoch because it predicts continued
spectral hardening at the highest Eddington ratios while the observed spectra appear 
to {\it soften}. This suggests the onset of additional physics which 
softens the spectrum as the Eddington ratio nears unity.

\acknowledgements{We gratefully acknowledge Ivan Hubeny for his instrumental contributions
to the development of BHSPEC.  We thank Eric Agol for useful discussions,
and for making KERRTRANS publicly available.  We also thank Marek Gierlinski,
Julian Krolik, Ramesh Narayan, Rebecca Shafee, and Aristotle Socrates for useful
discussions and/or assistance. We are grateful for comments on the manuscript from the 
anonymous referee and Jeff McClintock which lead to significant improvements.  Part of this
work was completed while the authors were hosted by the Kavli Institute for Theoretical 
Physics at UCSB. This work was supported by NASA grant NAG5-13228, and by the National Science 
Foundation under grant PHY99-07949.
}

\appendix
\section{A Simple Estimate for the Spectral Hardening}
\label{est}

A major focus of this work is understanding if a standard accretion disk model
can reproduce the spectra of the thermal dominant state, including its variation
with luminosity.  This task is difficult because it involves a self-consistent
calculation of the vertical structure and radiative transfer in a number of annuli.
This requires the solution of many coupled non-linear differential equations, and
sophisticated methods (e.g. Hubeny \& Lanz 1995) are needed to solve
this problem with precision.  However, it is insightful to first examine the extent
to which the results of the detailed calculation can be inferred from simple 
(albeit somewhat crude) arguments before comparing with the data.

We focus on a single annulus (the one with the largest $\teff$) which remains at
a fixed radius $R$ in the disk as $\ell$ varies. We make the
further approximation that
all the spectral variation is encapsulated in a single parameter $f$, the color
correction in equation \ref{e:ccbb}.  In the standard model $\ell=L/L_{\rm Edd} = 
\eta \dot{M}/\dot{M}_{\rm Edd}$, with $L_{\rm Edd}$, $\dot{M}_{\rm Edd}$, and 
$\eta$ fixed as $\mdot$ varies.  We can relate $\mdot$ to $\teff$ via
\be
\sigma \teff^4 = \frac{3G M \mdot}{8\pi R^3}
\label{e:flux}
\ee
where $\sigma$ is the Stefan-Boltzmann constant, $G$ is the gravitational constant,
and $\Omega$ Keplerian frequency. For simplicity we have dropped the `correction 
factors' due to relativistic effects and the no torque inner boundary condition.
Now we have reduced the problem of calculating the $L-T$ relation to finding
$f$ as a function of $\teff$.

In order to evaluate $f$ we need an approximate scheme for the spectral formation. In our
models, the spectra harden because electron scattering dominates the opacity, leading to deviations
from local thermodynamic equilibrium (LTE) and modified blackbody spectra.  The resulting spectra
are harder than a blackbody at $\teff$ as typical photon energies are higher.  The photons
are thermally emitted deeper in the atmosphere and their energy distributions are
characteristic of the temperatures where they are formed.  If the opacities are frequency
dependent the depth of formation (thermalization surface) varies with frequency and photons
at different frequencies originate from different depths with different temperatures.  However,
over a  broad range of temperatures
where bound-free processes dominate, the true absorption opacity is a much weaker
function of frequency than in the free-free case.  This `grey' opacity dependence yields
a spectrum which can be crudely approximated by a diluted Planck function with a 
temperature $\tfor$ evaluated at the depth of formation $\taufor$ (Davis et al. 2005).
Therefore, a simple estimate for $f$ is given by $f\sim\tfor/\teff$.  If the opacity
has a stronger frequency dependence, the Planckian shape will be a poorer approximation,
but this estimate of $f$ will still be crudely correct if $\taufor$ is evaluated
near the spectral peak.

In the electron scattering dominated atmospheres we are considering,
$\taufor \simeq m_\ast \kappaes$ where $\kappaes$ is the approximately constant
electron scattering opacity and
$m_\ast$ is the column mass where the effective optical depth is equal to unity.  It can be 
approximated by $m_\ast=1/\sqrt{3 \kappaes \kappaab}$, where $\kappaab$ is the absorption
opacity evaluated at $m_\ast$.  In general, $m_\ast$ is a frequency dependent quantity and
should be evaluated by integrating from the surface inward.  For our approximate estimates,
it is useful to ignore the frequency dependence and consider all quantities with an asterisk
subscript as being evaluated at a frequency near the spectral peak.  For simplicity, we assume an
absorption opacity of the form $\kappaab=\kappa_0 \rhofor \tfor^{-n}$ and find
\be
\taufor \simeq \left(\frac{\kappaes}{3\kappa_0} \right)^{1/2} \rhofor^{-1/2} \tfor^{n/2}.
\label{e:taustar1}
\ee

The temperature $T$ at an optical depth $\tau$ in an annulus can be estimated using an
LTE-grey model with mean opacities (Hubeny 1990), yielding
\be
T^4=\frac{3}{4}\teff^4 \left[\left(\tau - \frac{\tau^2}{\tau_{\rm tot}} + 3^{-1/2} \right)
+ \frac{2}{3 \Sigma \kappa_B} \right].
\label{e:temp}
\ee
Here, $\tau_{\rm tot}$ is the optical depth at the disk midplane,
and $\kappa_B$ is the Planck mean opacity.  We can use this expression to evaluate
the temperature $\tfor$ at the depth of formation $\taufor$
\be
\tfor^4\simeq \frac{3}{4}\teff^4 \taufor
\label{e:tstar}
\ee
where we have assumed $\taufor \ll \tau_{\rm tot}$, $\taufor >  3^{-1/2}$, and 
$\Sigma \gg \kappa_B^{-1}$. These limits are appropriate for annuli with large
$\Sigma$ and correspond to lower values of $\mdot$  and/or $\alpha$.
In this approximation, equation (\ref{e:tstar}) shows that $f$ depends only on
$\taufor$ through
\be
f \sim \frac{\tfor}{\teff}=\left(\frac{3}{4}\taufor\right)^{1/4}.
\label{e:fcol}
\ee

Next, we need an expression for density.  In the range of $\teff$, $\Sigma$, and
$\Omega$ of primary interest to us, the annuli are radiation pressure dominated at
the midplane. However, strong gas pressure gradients are still needed to support the
atmosphere near the surface (Hubeny 1990) where the radiation force
ceases to increase with height above the midplane $z$ as fast as the tidal gravity.
Hydrostatic equilibrium is then given by
\be
\cgas^2\frac{d \rho}{d m} \simeq \Omega^2 (z-z_0)
\label{e:he}
\ee
where $z_0$ is the height above which the gas pressure gradient dominates (of order
the radiation pressure scale height),
$\cgas^2=\kbol \tfor/(\mu m_{\rm p})$ is the local isothermal sound speed, $\mu$ is
the mean molecular weight, and $m_{\rm p}$ is the proton mass.  Using the definition
of column mass $d m= - \rho dz$, equation (\ref{e:he}) becomes
\be
\frac{\cgas^2}{\Omega^2}\frac{d \rho}{\rho} \simeq - (z-z_0) dz.
\label{e:rhoz}
\ee
We can use this to solve for $\rho$, approximating the atmosphere as isothermal. We find
\be
\rho=\rho_0 \exp{\left(-\frac{(z-z_0)^2\Omega^2}{2 \cgas^2}\right)}.
\label{e:rhodef}
\ee
This expression can be integrated to find $m_\ast$,
\be
m_\ast &=& \rho_0 \int^\infty_{z_\ast} \exp{\left(-\frac{(z-z_0)^2\Omega^2}{2 \cgas^2}\right)}
\nonumber \\
&=& \frac{\rho_0 \cgas}{\Omega}\sqrt{\frac{\pi}{2}} \,\,{\rm erfc}\left(\frac{\Delta z 
\Omega}{\sqrt{2} \cgas}\right)
\label{e:mstar}
\ee
where $\Delta z=z_\ast-z_0$ and erfc denotes the complimentary error function.  The argument
of the erfc is typically large in the range of interest.  In the limit
of large $x$, ${\rm erfc}(x)\simeq \exp(-x^2)/(\sqrt{\pi}x)$ and equation (\ref{e:mstar}) can
be evaluated to find
\be
\taufor=\kappaes m_\ast \simeq \frac{\kappaes \rhofor \cgas^2}{\Omega^2 \Delta z}.
\label{e:taustar2}
\ee
The expression for $\Delta z$ is difficult to approximate, and we must appeal to the full
atmosphere calculations for guidance.  Typically, both $z_\ast$ and $z_0$ increase with $\teff$
but their difference tends to {\it decrease}. It decreases only weakly in annuli where $\taufor \lesssim 100$
and $\Sigma \gtrsim 10^4$ g cm$^{-2}$, but more rapidly with $\teff$ in annuli 
where $\taufor$ is larger and/or $\Sigma$ is smaller. However, in these annuli the assumptions underlying
equation (\ref{e:taustar2}) are increasing invalid, so over the range of interest we approximate
$\Delta z$ as a constant.  Including the decrease in $\Delta z$ would make $f$ a stronger
function of $\teff$, but only weakly since $\taufor \propto \Delta z^{-1}$ and $f \propto \taufor^{1/4}$.
Since $\Sigma \propto \alpha^{-1}$, this approximation is better for lower $\alpha$.

Combining equations (\ref{e:taustar1}), (\ref{e:tstar}), and (\ref{e:taustar2}), we can solve
for the dependence of $\tfor$, $\taufor$, and $\rhofor$ on $\teff$:
\be
\frac{d \ln \tfor}{d \ln \teff} & = & \frac{12}{11-n} \nonumber\\
\frac{d \ln \rhofor}{d \ln \teff} & = & \frac{4n-8}{11-n} \nonumber\\
\frac{d \ln \taufor}{d \ln \teff} & = & \frac{4n+4}{11-n} \nonumber
\ee
where $n=3.5$ would be the relevant value for a Kramer's opacity law.  The bound-free
opacity dominates the true absorption opacity in the models at the temperatures of interest and
the mean opacity is not well approximated by a Kramer's law at these temperatures and densities.
We find empirically that a slightly weaker temperature dependence ($n\sim 2$) yields better 
agreement between equation (\ref{e:taustar1}) and the simulation results.  For $n=2$ we find
from equation (\ref{e:fcol}) that $f \propto \teff^{1/4} \propto \tfor^{1/3}$, so that we 
expect $f$ to increase with temperature, but with a weak dependence. The resulting $L-T$ 
relation is plotted as a dotted line in Figure \ref{fig4}, where we have used equation 
(\ref{e:flux}) to obtain $L/L_{\rm Edd} \propto \tfor^3$.  The normalization is arbitrary
and chosen to facilitate comparison with the models.  

It is worth noting the role that density gradients near the surface play in producing 
this trend. The choice of $n=2$ yields $\rhofor$ independent of $\teff$, a reasonably 
good approximation for model annuli with large $\Sigma$.
If we instead assume (as is commonly done) that radiation pressure dominated annuli
are constant density with $\rhofor$ equal to the midplane value of the density 
$\rho_{\rm mid}$, we find a much stronger density dependence since $\rho_{\rm mid} 
\propto \mdot^{-2} \propto
\teff^{-8}$ (Shakura \& Sunyaev 1973).  This produces a much stronger dependence of
$\taufor$ and $f$ on $\teff$.  It also yields a stronger dependence of $f$ on $\alpha$ since
$\rho_{\rm mid} \propto \alpha^{-1}$. The $\rhofor=\rho_{\rm mid}$ approximation becomes 
increasingly relevant when $\Sigma$ declines, either due to an increase in $\mdot$
or an increase in $\alpha$ (see e.g. Shakura \& Sunyaev 1973).  However, the large $f$'s that
would result are not typically seen in the data.  As pointed out by Shimura \& Takahara (1995),
this is largely the result of Compton scattering effects when $\tfor$ and $\taufor$ become
large.

It is difficult to address the effects of Compton scattering in detail without attempting
a full solution of the radiative transfer equation.  However, some insight can be gained by 
examining the behavior of the $y$-parameter (see e.g. Rybicki \& Lightman 1979)
\be
y = \frac{4 \kbol T- h \nu}{m_e c^2} N
\ee
where $T$ is temperature, $h$ is Planck's constant, $m_e$ is the mass of the electron,
$c$ is the speed of light, and $N$ is the number of electron scatterings.  The 
typical ratio
of final to initial photon energy is $E_{\rm f}/E_{\rm i} \sim \exp(y)$. In an annulus, the
temperature increases with optical depth as in equation (\ref{e:temp}).
Photons produced near $\taufor$ will tend to have higher energies than the
electrons they encounter as they scatter out of the atmosphere.  Therefore, $h \nu > 4 \kbol T$
so that the final photon energy is less than the initial photon energy and $y$ is negative.

For photons produced near $\taufor$, $h \nu \sim 4 \kbol \tfor$ and $N \sim \taufor^2$
for a photon reaching the photosphere.  Following Hubeny et al. (2001), we therefore 
define the `effective y-parameter' as
\be
y_\ast \equiv -\frac{4 \kbol \tfor}{m_e c^2}\taufor^2.
\label{e:ystar1}
\ee
The effects of Comptonization decrease the photon energies and bring them closer to LTE
near the photosphere, reducing the spectral hardening and lowering
$f$.  Instead of increasing indefinitely with increasing $\tfor$, $f$ will `saturate' when 
$E_{\rm f}/E_{\rm i} \simeq 1/f$ or $y_\ast \simeq -\ln f \sim -1$.  Using equation (\ref{e:fcol}),
we can rewrite equation (\ref{e:ystar1}) as
\be
y_\ast \simeq -\frac{\kbol \teff}{\rm 72 \,\, keV} f^9.
\label{e:ystar2}
\ee
For $y_\ast \sim -1$ and $\kbol \teff \sim 1$ keV, we find $f \sim 1.6$, close to the commonly
assume value of 1.7 (Shimura \& Takahara 1995).  Above this value, $f$ will increase only
weakly because further increase in the typical
energies of photons emitted at $\taufor$ is offset by reductions due to the effects
of Compton scattering.  The precise value of $f$ 
obtained will, of course, depend on the details, but the strong $f$ dependence in equation
(\ref{e:ystar2}) demonstrates why the spectra of effectively optically thick annuli are never consistent
with $f \gtrsim 2$.

We emphasize that the arguments discussed above are only approximate with varying accuracy 
over the parameter space of interest. They are not a substitute for full atmosphere calculations.
Nevertheless, comparison with these calculations (see e.g. Figure \ref{fig5})
demonstrates that they are qualitatively correct and are useful for understanding
why the spectral hardening depends so weakly on $\teff$ in most BHBs.

\end{document}